\documentclass[12pt,a4paper,onecolumn]{IEEEtran}

\usepackage{amsopn,amsmath,amssymb,amsfonts,bm,amsthm}
\usepackage{graphicx,subfigure}
\usepackage{enumerate}
\usepackage{cite}
\usepackage{etex}
\usepackage{algorithm,algorithmic}
\usepackage{tikz,pgfplots}
\usepackage{setspace}
\usepackage[margin=16mm,top=20mm,bottom=16mm]{geometry}
\newcommand{\figref}[1]{Fig.\,\ref{#1}}

\def\mA{\boldsymbol A}
\def\E{\mathcal E}
\def\e{\boldsymbol e}

\def\F{\mathbb F}
\def\mG{\mathcal G}
\def\g{\gamma}
\def\G{\boldsymbol G}
\def\H{\mathcal H}
\def\O{\mathcal O}
\def\P{\mathcal{P}}
\def\p{\mathbf{p}}
\def\R{\mathcal{R}}
\def\r{r}

\def\V{\mathcal V}
\def\v{\boldsymbol v}

\newtheorem{Theorem}{\textbf{Theorem}}

  {\proof}{\proofend}

\newtheorem*{problem*}{\textbf{Problem MP}}

\title{The Benefit of Limited Feedback to {Generation-Based} Random Linear Network Coding in Wireless Broadcast}
\author{Mingchao Yu*, Parastoo Sadeghi*, and Alex Sprintson$^\dag$ \\ \small{*Research School of Engineering, Australian National University, Canberra, Australia\\$^\dag$Department of Electrical and Computer Engineering, Texas A\&M University, Texas, USA\\ \texttt{Emails: \{ming.yu,parastoo.sadeghi\}@anu.edu.au,~spalex@tamu.edu}}}
\begin{document}
\maketitle
\vspace{-4.5em}
\begin{abstract}
Random linear network coding (RLNC) is asymptotically throughput optimal in the wireless broadcast of a block of packets from a sender to a set of receivers, but suffers from heavy computational load and packet decoding delay. To mitigate this problem while maintaining good throughput, we partition the packet block into disjoint generations after broadcasting the packets uncoded once and collecting one round of feedback about receivers' packet reception state. We prove the NP-hardness of the optimal partitioning problem by using a hypergraph coloring approach, and develop an efficient heuristic algorithm for its solution. Simulations show that our algorithm outperforms existing solutions.
\end{abstract}

\section{Introduction}

In many wireless settings there is a need to broadcast a block of data packets from one sender to multiple receivers over independent erasure channels \cite{keller:drinea:fragouli:2008}. A key problem in this context is to find an efficient broadcast scheme that allows all receivers to obtain all the data packets in the presence of erasures. It has been shown that the \emph{random linear network coding (RLNC)} \cite{ho:medard:koetter:karger:effros:2006} technique can almost surely minimize the number of transmissions, and, in turn, maximize throughput by transmitting coded packets, which are random linear combinations of all the data packets using a sufficiently large finite field. RLNC relies on \emph{block-decoding}, i.e., all data packets can only be decoded when a sufficient number of coded packets have been received. However, this approach results in large packet decoding delays \cite{keller:drinea:fragouli:2008} and high computational load \cite{heide2011code}. One possible way to reduce computational load is to use the sparse RLNC technique \cite{fragouli:widmer:boudec:2008}, in which the linear coefficients are more likely to be zeros. However, its packet decodings are still based on the whole block, hence the packet decoding delay is still large.

In this letter, we focus on RLNC-based broadcast schemes that partition the block of data packets into \emph{generations} and construct the coded combinations from data packets that belong to the same generation.
This technique results in a substantial reduction of  the computational load. It also allows decoding of data packets that belong to the current generation before the next generation is transmitted, resulting in a significant decrease in the packet decoding delay.

While generation-based RLNC techniques are efficient in reducing the packet decoding delay and minimizing the computational load, they result in a larger number of transmissions, and, as a result, lower throughput. The main problem in this context is to design throughput-efficient generation-based RLNC schemes.  This issue has been the subject of many studies, e.g.,  \cite{maymounkov:generation:2006,silva:generation:2009,heidarzadeh:overlapped:2010,emina:li:isit2012,emina:joshi:isit2013}. Most of the previous studies, however, focused on settings with no receiver feedback. In such settings, the sender is unaware of what packets are missing at each receiver, hence it cannot utilize this information to optimize the transmission scheme.

To the best of our knowledge, only few studies have focused on the transmission schemes with feedback. Yu et al. \cite{yu:parastoo:neda:2014}  considered the partitioning problem based on solutions borrowed from instantly decodable network coding (IDNC).
However, since IDNC requires instant decodability of the coded combinations, it does not provide insight into the partitioning problem when more general coding techniques are used.  Reference \cite{rezaee2011smart} proposes a limited feedback scheme that utilizes RLNC with block decoding. This scheme, however, still suffers from heavy computational load and large packet decoding delays.

In this letter, we study the general partitioning problem by considering a two-phase broadcast scheme with limited receiver feedback. In the first phase, each data packet in the block is transmitted once in an uncoded form. Then, the receivers feed back their packet reception state to the sender. Based on this information, the sender partitions the packet block into generations and broadcasts a set of coded packets, where each coded packet is a combination of packet that belong to the same generation. The receivers use coded packets to recover missing data packets.

Our contributions can be summarized as follows: (i) We formulate an optimal partitioning problem and establish its NP-hardness; (ii) We propose a heuristic partitioning algorithm that strikes a balance between decoding delay and block completion time and provides a local Pareto-optimal solution.
(iii) We perform an extensive simulation study that shows that our algorithm can significantly outperform the previously proposed methods in the literature.

\section{Problem Formulation}
\begin{figure}
\centering
\subfigure[State feedback matrix $\mA$]{\includegraphics[width=0.3\linewidth]{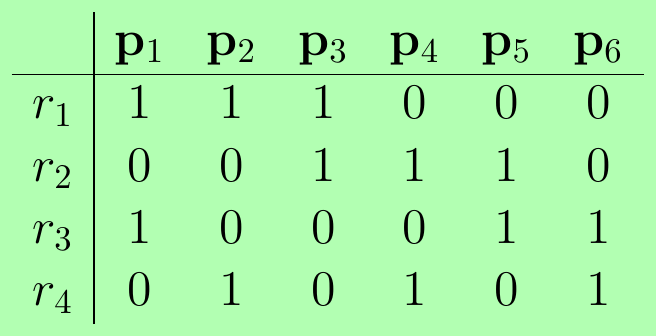}}\hspace{30pt}
\subfigure[Hypergraph  $\H$]{\includegraphics[width=0.2\linewidth]{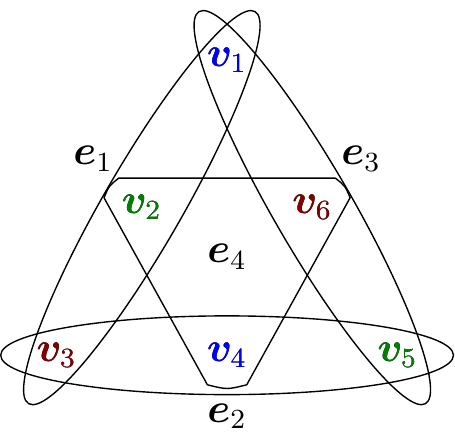}}
\caption{A state feedback matrix and its hypergraph representation. The minimum 1-regular coloring solution of $\H$ requires 3 colors.}
\label{fig:sfm_graph}
\end{figure}

We consider a scenario in which a sender wants to broadcast a block of $K$ data packets $\P=\{\p_k\}_{k=1}^K$ to a set of $N$ receivers $\R=\{\r_n\}_{n=1}^N$ through wireless channels which are subject to independent random packet erasures.
In the first phase, referred to as the \emph{systematic transmission phase}, the sender broadcasts each packet in $\P$  without coding. Then, the sender collects one round of feedback from each receiver about its packet reception state.  This information is represented by a binary $N\times K$ state feedback matrix (SFM) $\mA$, where $\mA(n,k)=1$ means that $\r_n$ has missed $\p_k$ and thus still wants it, and $\mA(n,k)=0$ means $\r_n$ has received $\p_k$. We denote by $T(k)$ the number of receivers that want $\p_k$. An example of SFM with $T(1)=2$ is given in \figref{fig:sfm_graph}(a).

Based on the SFM, the sender then partitions $\P$ into a set of  $M$ disjoint \emph{generations}, denoted by $\G=\{\mG_m\}_{m=1}^M$.
We define the \emph{rank}  $\g_m$ of a generation $\mG_m$ as the maximum number of packets in $\mG_m$ wanted by one receiver, i.e.,

\vspace{-1em}
\begin{equation}
\g_m=\max_{\r_n\in R}\sum_{\p_k\in \mG_m}\mA(n,k).
\end{equation}
For example, for the SFM in \figref{fig:sfm_graph}(a), generation $\{\p_2,\p_3\}$ has a rank of 2.

The sender then starts a \emph{coded transmission phase}. In each transmission, it broadcasts a coded packet of a generation, which is a linear combination of data packets from this generation. We consider random linear network coding (RLNC) \cite{ho:medard:koetter:karger:effros:2006}, which chooses the linear coefficients uniformly at random from a finite field $\F_q$. When $\F_q$ is sufficiently large, RLNC asymptotically guarantees linear independency between coded packets, so that any receiver who wants $\gamma_m$ packets from $\G_m$ can decode these packets after receiving $\gamma_m$ coded packets of $\G_m$ with high probability. Note that this decoding condition can also be achieved by deterministic codes such as maximum-distance-separation (MDS) codes and fountain codes \cite{shokrollahi2006raptor,luby2002lt}. We focus on RLNC for clarity, but our techniques can be extended to other codes.

In our context, the original RLNC technique uses only one generation, i.e., $\G_1=\P$. When it is applied, a receiver who is missing $\g_1$ packets from $\P$ needs to receive $\g_1$ linearly independent coded packets of $\P$ before being able to decode them, which incurs a computational complexity of $\O(\g_1^3)$ \cite{heide_systematic_RLNC}. To minimize the computational load and reduce the packet decoding delay, while also maintaining a good throughput performance,  we propose the following \emph{minimum partitioning} (MP) problem:

\begin{problem*}
Given an SFM $\mA$ and a constant $\g$, find the minimum value of $M$, for which there exists a partition of the packet block $\P$ into $M$ disjoint generations $\G=\{\mG_m\}_{m=1}^M$ such that the rank of each generation $\mG_m$ does not exceed $\g$, i.e., $\g_m\leq \g$ for $m\in[1,M]$.
\end{problem*}

Minimizing the total number of generations  $M$ is an important goal due to the following reasons:

\begin{enumerate}
	\item The lower values of $M$ would result in smaller number of transmissions and, as a result, higher throughput. In particular, we need at least $U\triangleq\sum_{m=1}^M\g_m \le M\g$  transmissions in the coded transmission phase to satisfy all receivers. Due to erasures, more transmissions may be needed, but its number will depend on $M$, and lower values of $M$ typically require a smaller number of transmissions.

\item The lower values of $M$ reduce the overall  computational load of decoding algorithm. Indeed, every receiver can decode the data packets it wants from any generation by solving a set of at most $\g$ linear equations. This bounds the total computational load of decoding to $\O(M\g^3)$ operations.
\item The lower values of $M$ reduce the average packet decoding delay (APDD) $D$, i.e., the average time it takes for a receiver to decode a data packet. It is calculated as:
    \vspace{-1em}
	\begin{equation}\label{eq:d_def}
	D=\frac{1}{\sum_{k=1}^KT(k)}\sum_{\mA(n,k)=1}u_{n,k},
	\end{equation}
	where $u_{n,k}$ is the time index in the coded transmission phase when $\r_n$ decodes $\p_k$.

Also note that the APDD of a partition is upper bounded as  \cite{yu:parastoo:neda:2014}:
\begin{equation}
D \leqslant\sum_{m=1}^M\frac{\g_m^2+\g_m}{2},
\end{equation}
so that the lower values of $M$ typically result in lower values of $D$.
	
\end{enumerate}

\section{The Hardness of the Minimum Partitioning Problem}
In this section, we analyze the hardness of Problem MP by constructing a reduction from a hypergraph coloring problem. A hypergraph $\H$ is defined by a pair $(\V,\E)$, where $\V$ is the set of vertices, and $\E$ is the set of hyperedges. Every hyperedge $\e\in\E$ is a subset of $\V$ with size $|\e|\geqslant 1$. A hypergraph is $\omega-$uniform if for every hyperedge $\e\in\E$ it holds that $|\e|=\omega$. A $\g$-regular coloring of $\H$ with $M$ colors is a partition $\{\V_m\}_{m=1}^M$ of $\V$ that satisfies $|\V_m\cap\e_n|\leqslant\g$ for each $\e_n\in\E$. The minimum number $M$ of colors required for a $\g$-regular coloring of $\H$ is called the $\g$-chromatic number of $\H$. It was shown in \cite{krivelevich1998chromatic,krivelevich2003approximate} that the problem of finding the $\g$-chromatic number for a given hypergraph is NP-complete.

Our reduction includes the following steps. Given $\H$, for every vertex $\v_k\in\H$ we generate a data packet $\p_k$, and for every hyperedge $\e_n$ we generate a receiver $\r_n$ who wants the data packets that correspond to the vertices that are incident to $\e_n$. For example, the SFM in Fig. \ref{fig:sfm_graph}(a) can be generated from the hypergraph depicted in Fig.~\ref{fig:sfm_graph}(b), with 6 data packets/vertices and 4 receivers/hyperedges.

Let $\{\V_m\}_{m=1}^M$ be a $\g$-regular coloring solution of the instance $\H$ of the $\g$-regular coloring problem. Then, let $\G=\{\mG_m\}_{m=1}^M$  be a partition of $\P$ such that for $m=1,\dots,m$, a generation $\mG_m$ includes packets that correspond to vertices in $\V_m$. It is easy to verify that $\G$ is a valid solution to Problem MP with $M$ generations such that the rank of each generation is bounded by $\g$.

Similarly, for any solution $\G=\{\mG_m\}_{m=1}^M$ of Problem MP, we can construct a  $\g$-regular coloring solution $\{\V_m\}_{m=1}^M$ with $M$ colors, in which each set $\V_m$ includes vertices that correspond to packets in $\mG_m$. Thus, an optimal algorithm for Problem MP can be used to find the $\g-$chromatic number of a hypergraph, which is an NP-complete problem. We summarize our results in the following theorem:

\begin{Theorem}
Problem MP is NP-complete.
\end{Theorem}

We note that our reduction can also be used to establish the hardness of several special cases of our problem based on other results in graph theory. For example, it is an NP-complete problem to find 2-regular coloring with two colors for a $3$-uniform hypergraph \cite{Lovasz:color,krivelevich2003approximate}. This implies that, in the special case in which every receiver wants $\omega=3$ data packets, it is NP-complete to find a partition of size $M=2$ with the rank of each generation being bounded by $\g=2$. In addition, it is NP-complete to find the chromatic number of hypergraphs in the special case of 1-regular coloring \cite{agnarsson2005strong}.
This implies that our problem is intractable even for the special case of $\g=1$, in which every generation can include at most one missing packet for any receiver \cite{yu:parastoo:neda:2014}.
Thus, by using our technique we can recover the hardness result of instantly decodable network coding established in \cite{Rozner_Heuristic_clique} using a different approach.
\section{Partition Algorithm}\label{sec:algorithm}

In this section, we propose a heuristic algorithm for Problem~MP.
In addition to aiming to minimizing the number of partitions, our algorithm also orders the partitions in such a way that  popular packets (i.e., packets with large values of $T(k)$) are assigned to early generations. The partition ordering satisfies the \emph{irreducibility} property, i.e., no data packet from a generation $\mG_m$ can be moved to an earlier generation $\mG_n$ ($n<m$) without increasing its rank \cite{parastoo:yu:neda:isita}. The irreducibility property, in turn, implies that the proposed algorithm achieves local \emph{Pareto-optimal} trade-off between throughput and packet decoding delay delay.

Note that another approach for Problem MP is to find the corresponding instance of the hypergraph coloring problem and then use existing hypergraph coloring algorithms. However, existing approximation algorithms for the hypergraph coloring problem mainly focus on the special case of uniform hypergraphs (e.g., \cite{krivelevich2003approximate}) and do not consider other aspects of the solution mentioned above.

The detailed description of the algorithm is provided in Algorithm~1. The algorithm receives as inputs a state feedback matrix $\mA$ and the upper bound $\g$ of the rank of each generation. The algorithm iteratively creates generations and fills them with packets. At each step, the algorithm searches for a packet that can be added to the current generation without increasing its rank. If such a packet exits, it is added to the generation. If such a packet does not exits and the rank of the current generation is less than $\g$, then the most popular packet  in $\P$ is added to the current generation. A packet that is added to the current generation is removed from $\P$. If the the size of the current generation is $\g$ and no packet can be added to it without increasing its rank, than a new generation is formed. The algorithm starts with empty generation $\mG_1$ and terminates when set $\P$ becomes empty.

\begin{algorithm}[t]
\caption{Feedback-assisted minimum partition}
\label{alg:partition}
\begin{algorithmic}[0.8]
\STATE Initialize: SFM $\mA$, generation rank $\g$, a counter $m=0$ 

\WHILE{$\P$ is nonempty}
\STATE $m\leftarrow m+1$, create an empty generation $\mG_m$;
\WHILE{there exists a packet in $\P$ that would not increase the current rank of $\mG_m$, $\g_m$, to $\g+1$}
\IF {there exists a packet in $\P$ that would not increase $\g_m$ by one}
\STATE add the most popular packet that satisfies this condition to $\mG_m$, and then remove it from $\P$;
\ELSE
\STATE add the most popular packet in $\P$ to $\mG_m$, and then remove it from $\P$. Increase $\g_m$ by one;
\ENDIF
\ENDWHILE
\ENDWHILE
\end{algorithmic}
\end{algorithm}

\section{Numerical results}\label{sec:simulation}

\begin{figure}[b]
\centering
\includegraphics[width=0.5\linewidth]{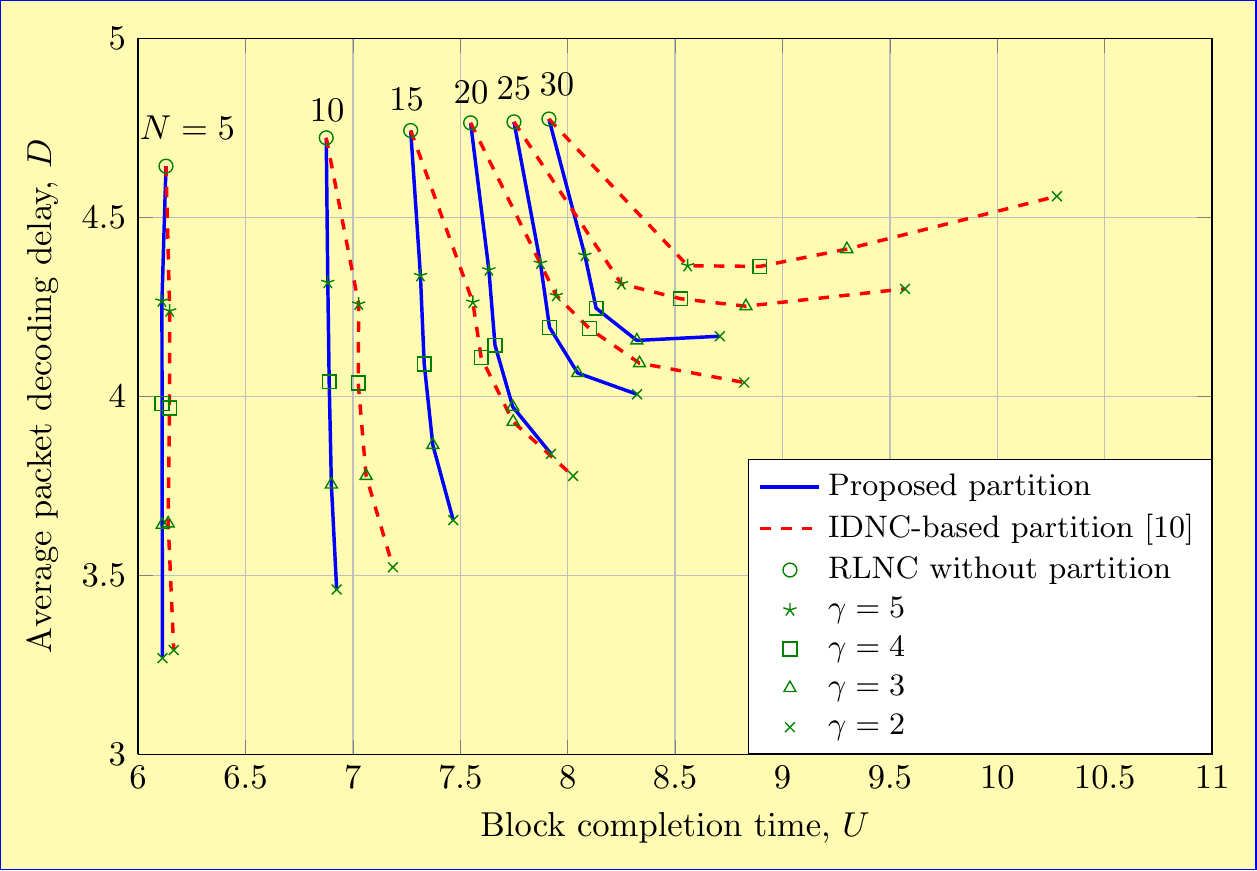}
\caption{The throughput-delay tradeoffs of the proposed and IDNC-based partition algorithms for different numbers of receivers $N$.}
\label{fig:tradeoff_g}
\end{figure}

In this section, we compare the performance of our algorithm with alternative solutions.
Note that once the packet block is partitioned into generations, there are several different ways to schedule transmissions in the coded transmission phase to satisfy all the receivers. In our experiments, we used the following strategy. The sender broadcasts the generations in a round-robin manner until all receivers are satisfied and feedback ACKs. In every transmission round $\g_m$ coded packets of each generation $\mG_m$ are broadcast. We note that several more advanced strategies may also be used. For example, the sender can send more than $\g_m$ coded packets of each $\mG_m$ in the first round to increase the probability of block completion after this round, or it can merge generations together with the help of extra feedback to reduce redundant packets \cite{rezaee2011smart,yu:parastoo:neda:2014}.

\figref{fig:tradeoff_g} compares the performance of the partition solutions produced by our algorithm with the IDNC-based algorithm presented in \cite{yu:parastoo:neda:2014} in terms of throughput-delay performance, measured by the block completion time $U$ and average packet decoding delay $D$, respectively. In our experiments, the packet block size is $K=20$ and the packet erasures in the systematic transmission phase are random i.i.d. random variables with a probability of $P_e=0.2$. In this figure only, we assume that the coded transmissions are erasure-free. Our experimental results show that the proposed algorithm always outperforms the IDNC-based one in terms of $U$. The improvement becomes more significant with increasing number of receivers. The $D$ of the two algorithms is similar when $\g$ is large, but our algorithm outperforms the IDNC-based one with increasing number of receivers when $\g$ is small.

\begin{figure}[t]
\centering
\subfigure[Block completion time]{\includegraphics[width=0.5\linewidth]{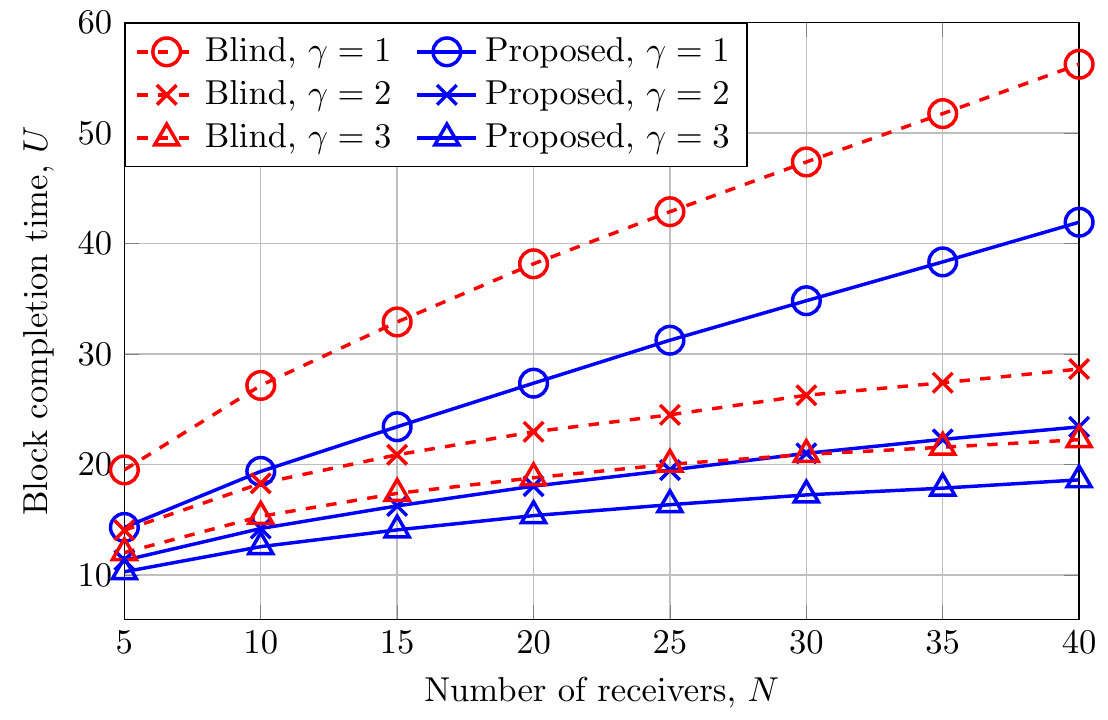}}
\subfigure[Average packet decoding delay]{\includegraphics[width=0.5\linewidth]{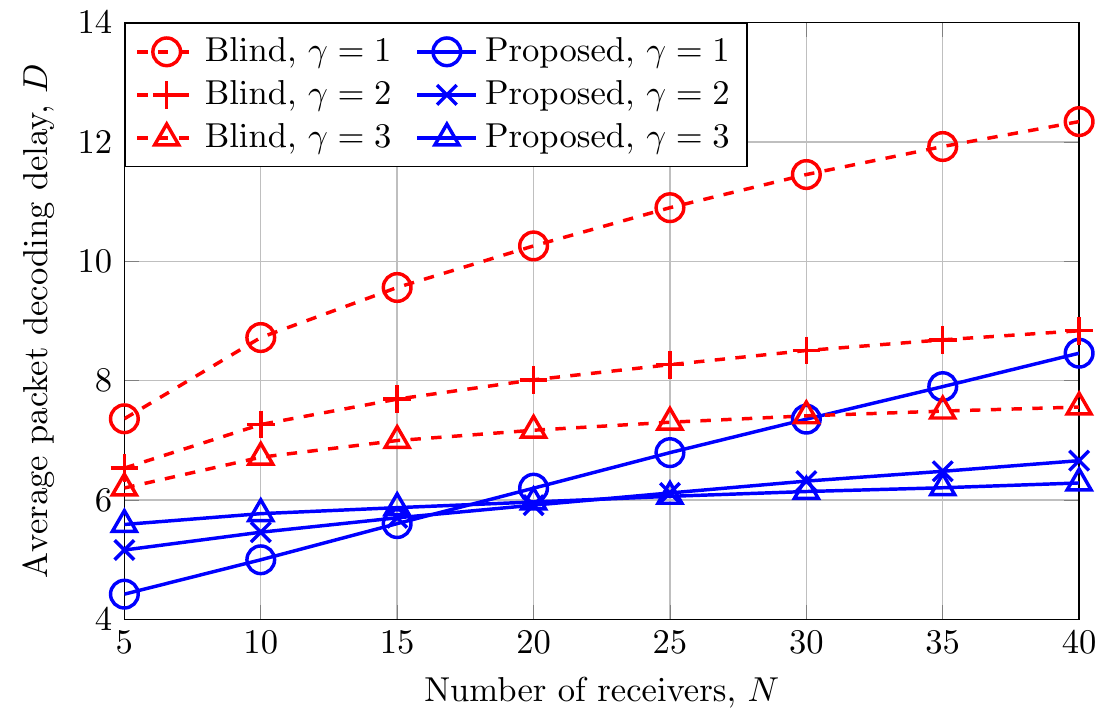}}
\caption{Performance improvements with one round of feedback.}
\label{fig:performance_gain}
\end{figure}

We also evaluate the gain achieved by our algorithm, enabled by the feedback collected after the systematic transmission phase. To this end, we compare the throughput and decoding delay of the coded transmission phase when our algorithm and the \emph{blind} partition approach are applied, respectively. With the blind partition approach, we sequentially segment the packet block into $M$ generations, each carrying $\frac{K}{M}$ data packets, where $M$ is set to be the number of generation produced by our algorithm. If the blind partition is applied, only one coded packet of every generation is broadcast in each round, as the sender is unaware of the generation ranks.

In our experiments,  the packet block size is $K=20$, and the packet erasures are random i.i.d. random variables with a probability of $P_e=0.2$ in both the systematic and coded transmission phases. The simulation results are shown in \figref{fig:performance_gain}. We observe that by collecting only one round of feedback, our algorithm can reduce the block completion time by 30\%  and the average packet decoding delay by 40\%, compared to the blind partition. The improvement decreases with increasing $\g$, because both techniques will converge to the original RLNC.
\section{Conclusion}
In this letter, we showed that feedback-assisted partitioning can bring a better tradeoff in terms of throughput, computational load, and packet decoding delay in RLNC-based wireless broadcast. We established the NP-completeness of optimal partition problem and presented a heuristic partition algorithm that achieves local Pareto-optimality. Our extensive simulations show that the proposed algorithm outperforms previously proposed solutions. As a future work, we are interested in developing  generation-based approaches that use deterministic coding techniques, such as  MDS codes and fountain codes.

We are also interested in designing a comprehensive solution to the wireless broadcast problem that includes partitioning, scheduling, and an optimal feedback mechanism.

We believe that the connection between wireless broadcast and hypergraph coloring established in this letter will provide new insights into other network settings such as cooperative data exchange problem.
\begin{spacing}{1.9}
\bibliographystyle{IEEEtran}

\end{spacing}
\end{document}